\documentclass[
    ,final            
  graphicx                 
  ]
  {aipproc}

\layoutstyle{6x9}

\begin{document}

\title{The luminosity function of cluster pulsars}

\classification{97.60.Gb, 98.20.Gm, 02.70.Uu}
\keywords      {pulsars: general -- methods: numerical -- methods: statistical -- globular clusters: general }

\author{Manjari Bagchi}{
  address={Department of Physics, Hodges Hall, West Virginia University, Morgantown, WV 26505, USA}
}

\author{Duncan R. Lorimer}{
  address={Department of Physics, Hodges Hall, West Virginia University, Morgantown, WV 26505, USA},
  altaddress={NRAO, Green Bank Observatory, PO Box 2, Green Bank, WV 24944, USA}
}

\begin{abstract}

We study luminosities of millisecond pulsars in globular clusters by fitting the observed luminosity distribution with single and double power laws. We use simulations to model the observed distribution as the brighter part of some parent distribution for Terzan 5 and try to find a model which simultaneously agrees with the observed diffuse radio flux, total predicted number of pulsars and observed luminosity distribution. We find that wide ranges of parameters for log-normal and power-law distributions give such good models. No clear difference between the luminosity distributions of millisecond pulsars in globular clusters and normal disk pulsars was seen.

\end{abstract}

\maketitle


\section{Introduction}

Millisecond pulsars (MSPs) in globular clusters (GCs) can be used as a tool to understand the properties of GCs as well as the recycling process in the dense stellar environments inside GCs \citep{cr05}. As luminosity is a fundamental property of pulsars, one necessary step to achieve this goal is to understand the luminosity distribution of MSPs in GCs. A full dynamical approach, where one models evolution of pulsars and observational limits following appropriate choice of birth distributions of pulsar parameters, can not be adopted for GC pulsars where it is difficult to model the effects of stellar encounters and the cluster potential. The simplest way is to use a snapshot approach where one models pulsar luminosities as observed. We first adopt this method and then, using Monte Carlo simulations, try to find a good model which not only fits the observed luminosity distribution, but also agrees with the total radio and $\gamma$-ray fluxes for Terzan 5.

\section{Analysis and results}

\begin{figure}[h]
\begin{tabular}{rl}
\includegraphics[width=0.5\textwidth,height=0.25\textheight]{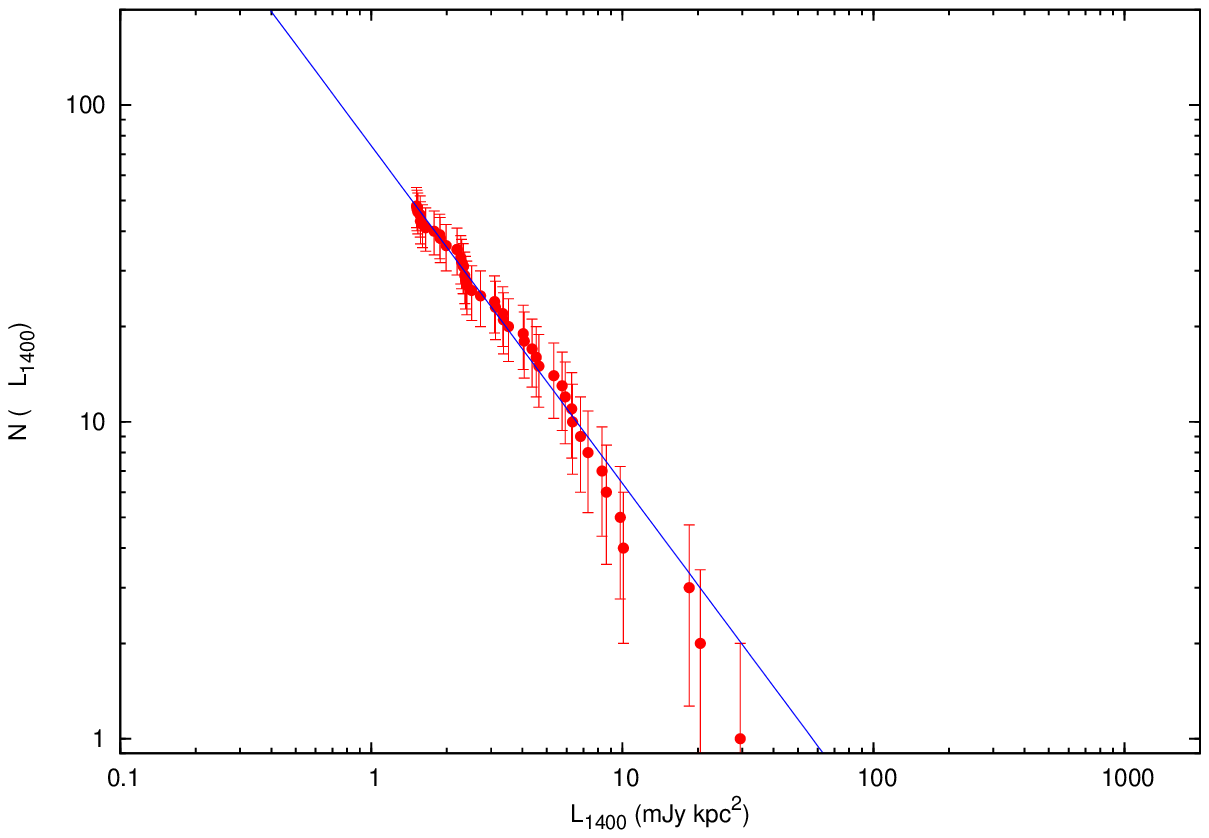}
\includegraphics[width=0.5\textwidth,height=0.25\textheight]{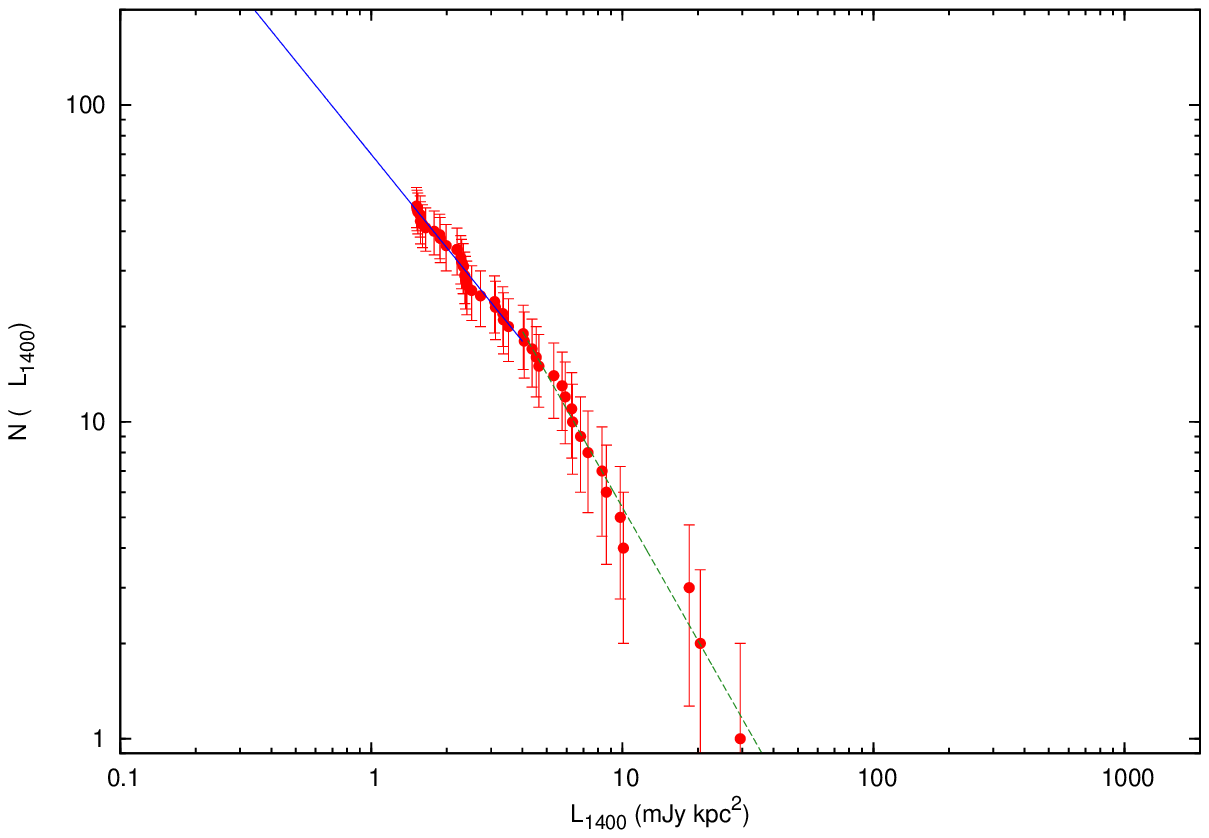}
\end{tabular}
\caption{Single (left) and double power-law (right) fit of GC pulsars with $L_{1400} > 1.5~{\rm mJy~kpc^2}$. Double power-law fit parameters as $N_{0,l}=70^{+7}_{-6}$, $q_{l}= -0.97 \pm 0.13 $ for $L_{1400} \leq 4.0 {\rm ~mJy~ kpc^2}$ and $N_{0,h}=134^{+60}_{-41}$, $q_{h}=  -1.40 \pm 0.21 $ for $L_{1400} > 4.0 {\rm ~mJy~ kpc^2}$.}
  \label{fig:simpleGC}
\end{figure}

We consider GC pulsars with known flux values and spin periods smaller than 100 ms and exclude the GCs having less than four such pulsars. Thus our dataset contains 79 pulsars in 9 GCs. We use the flux values measured at 1400 MHz ($S_{1400}$) if available, otherwise we estimate the value of $S_{1400}$ setting the value of the spectral index ($\alpha$) to be $-1.9$, which is the mean value obtained from the MSPs in GCs having flux values reported at multiple frequencies. This value agrees with the earlier estimate for 19 MSPs including two GC MSPs \cite{tbms98}. ``Pseudo-luminosities'' ($L_{1400}$) of the pulsars have been calculated using the relation $L_{1400} = S_{1400} \, d^2$ where $d$ is the distance of the host GCs from the sun. We confirm the earlier conclusion \cite{hrskf07} that the choice of $\alpha$ in a realistic range does not significantly affect the complementary cumulative distribution (CCD) of pulsar luminosities. With these data, we first revisit the study of Hui et al. (2010; hereafter HCT10) \cite{hct10} who modeled CCDs of luminosities of GC pulsars as $N(\geq L_{1400}) = N_0 \, L_{1400}^{q}$. We find $N_0=59 \pm 1$, $q=-0.80 \pm 0.03$ when we fit all 79 pulsars and $N_0=74^{+5}_{-4}$, $q= -1.06 \pm 0.06 $ when we fit 48 pulsars with $L_{1400} > 1.5~{\rm mJy~kpc^2}$; whereas HCT10 found $N_0=68 \pm 2$, $q=-0.58 \pm 0.03$ when they fitted all 78 pulsars and $N_0=91 \pm 6$, $q= -0.83 \pm 0.05 $ when they fitted 58 pulsars with $L_{1400} > 1.5~{\rm mJy~kpc^2}$. The fit improves if we keep only pulsars with $L_{1400} > 1.5~{\rm mJy~kpc^2}$. Moreover, we find that the fit significantly improves if we adopt two power laws. Fig. \ref{fig:simpleGC} shows the single and double power law fit of GC pulsars with $L_{1400} > 1.5~{\rm mJy~kpc^2}$. We also fit a single power law for pulsars with $L_{1400} > 0.5~{\rm mJy~kpc^2}$ in individual GCs separately and find that our fitting parameters are different from those obtained by HCT10 shown in Table \ref{tab:simpleGC}. This is due to two reasons. Firstly, our dataset is somewhat different. HCT10 did not exclude pulsars with spin period greater than 100 ms, they kept the pulsars in NGC 6441 which we exclude, we have 7 pulsars in M15 but HCT10 did not consider this GC at all. Secondly, we use recent distance estimates of GCs which are usually different from those used by HCT10, the largest difference is for Terzan 5. 

\begin{table}[h]
\begin{tabular}{lrrrr}
\hline
  \tablehead{1}{l}{b}{GC \\ name}
  & \tablehead{2}{c}{b}{this work \\ $N_0$ ~~~~~~~~~~~~~~~~~~~~~~~ $q$}
  & \tablehead{2}{c}{b}{HCT10  \\$N_0$  ~~~~~~~~~~~~~~~~~~~~~~~ $q$}   \\
\hline
47Tuc & $10^{+1}_{-1}$ & $ -0.85 \pm 0.18 $  & $11^{+2}_{-2}$    & $ -0.82 \pm 0.19$ \\
M3 & $2^{+1}_{-1}$ & $ -1.52 \pm 1.14$ & $2^{+1}_{-1}$  & $ -1.61 \pm 1.09 $ \\
M5 & $3^{+1}_{-1}$ & $ -0.55 \pm 0.32$ & $3^{+1}_{-1}$  & $ -0.58 \pm 0.38$ \\
M13 & $4^{+1}_{-1}$ & $ -0.62 \pm 0.39 $ & $4^{+2}_{-1}$  & $ -0.63 \pm 0.34$ \\
Ter5 & $20^{+1}_{-1}$ & $ -0.87 \pm 0.10$ & $50^{+12}_{-9}$  & $ -0.80 \pm 0.12 $ \\
NGC 6440 & $11^{+12}_{-6}$ & $ 0.86 \pm 0.53$ & $10^{+7}_{-4}$  & $ -0.59 \pm 0.27 $ \\
NGC 6441 & -- & -- & $8^{+14}_{-5}$  & $ -0.76 \pm 0.52$ \\
M28 & $12^{+4}_{-3}$ & $ 0.91 \pm 0.31$ & $10^{+5}_{-4}$  & $-0.74 \pm 0.26 $ \\
NGC 6752 & $5^{+2}_{-1}$ & $ -0.78 \pm 0.44 $ & $5^{+2}_{-2}$  & $ -0.93 \pm 0.50 $ \\
M15 &  $8^{+3}_{-2}$& $ -0.83 \pm 0.34 $ & --  & -- \\
\hline
\end{tabular}
\caption{Power Law fit parameters for different GCs}
\label{tab:simpleGC}
\end{table}

\begin{figure}[h]
{\includegraphics[width=1.1\textwidth,height=0.7\textheight]{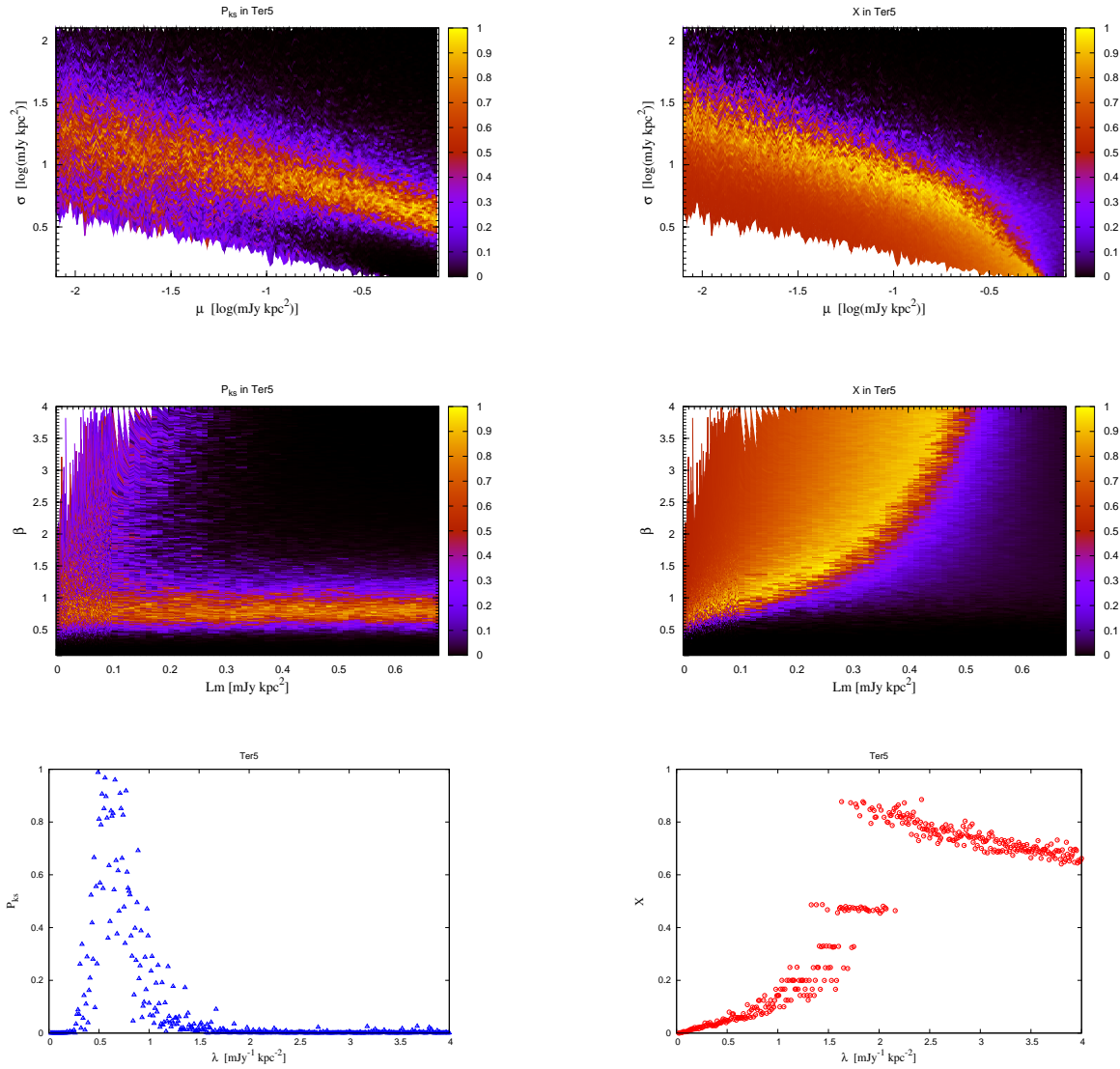}} 
\caption{Plot of $P_{ks}$ and $X$ for log-normal, power-law and exponential distributions. For log-normal and power-law distributions, we plot the parameters along the x and y axes and the value of $P_{ks}$ or $X$ as color codes. The parameters for the log-normal distribution are mean ($\mu$) and standard deviation ($\sigma$), and the parameters for the power-law distribution are minimum luminosity ($L_{m}$) and index ($\beta$). For the exponential distribution, we show the value of the parameter ($\lambda$) along the x axis and the value of $P_{ks}$ or $X$ along the y axis.} \label{fig:sim_all}
\end{figure}

Similarly, one can use any other distribution function to fit the observed CCD, but instead of doing so, we invoke a better method                                                       which simultaneously agrees with the observed diffuse radio flux ($S_{o,tot}$), total predicted number of pulsars ($N_{tot}$) and the observed
luminosity distribution. The luminosities of $N_{tot}$ pulsars are simulated from a chosen distribution and the distribution of luminosities of $N_{s}$ pulsars having $L \geq L_{m,o}$ is compared with the observed distribution using KS tests where the KS probability $P_{ks}$ should be high for a good model. $L_{m,o}$ is the observed minimum luminosity. But there can be models which give large/small values of $N_s$ in comparison to the observed number of pulsars ($N_{o}$), but the CCD having the same shape as the observed CCD. This can not be a good model although $P_{ks}$ will be large. To overcome this problem, we check whether the ``goodness factor'' $X = \frac{1}{1+Q}$ is $\sim 1$. Here we define
\begin{equation}
Q=\frac{(S_{s,tot}-S_{o,tot})^2}{(S_{o,tot})^2}+\frac{(N_{s}-N_{o})^2}{N_{o}^2},~~{\rm where} ~ S_{s,tot} = \displaystyle\sum\limits_{i=0}^{N_{tot}} S_{i}
\end{equation} 
for simulated fluxes $S_{i}$. Presently we apply this method only in the case of Terzan 5 for which $N_{tot}$ has been predicted to be $\sim 180$ from the total $\gamma$-ray flux \cite{abd10} and $S_{o,tot}$ was estimated to be 5.22 mJy \cite{fg00}. We use log-normal, power-law and exponential distribution functions over a wide range of parameters. We plot $P_{ks}$ and $X$ for all three distributions in Fig. \ref{fig:sim_all}. We plot the parameters of log-normal and power-law distributions along x and y axes and the values of $P_{ks}$ and $X$ as color codes. For the single-parameter exponential distribution, we show the value of the parameter (${\rm {mean}^{-1}}$) along the x axis and the value of $P_{ks}$ and $X$ along the y axis. For log-normal and power-law distributions, we find wide ranges of parameters where both $P_{ks}$ and $X$ have high values, so it is difficult to pinpoint good luminosity distribution functions. But for the exponential distribution we do not see a parameter space where both $P_{ks}$ and $X$ have high values and conclude that the exponential distribution can not describe pulsar luminosities adequately. 
In summary, we find that both log-normal and power-law distributions can reproduce the observed luminosities of MSPs in Terzan 5 but it is difficult to find exact parameters for a good model. Moreover, the model invoked \cite{fk06} for normal disk pulsars (FK06 model, log-normal distribution with mean -1.1 and standard deviation 0.9) falls in the good region (Fig. \ref{fig:sim_all}), suggesting that GC MSPs might have the same luminosity distribution as normal disk pulsars. It will be interesting to perform similar studies for other GCs which may constrain the distribution functions leading to the answer of the question whether GC pulsars form a different population from disk pulsars.

\bibliographystyle{aipproc}

\end{document}